\documentstyle[11pt,epsfig]{article}
[1999/03/29 PSNFSS v.7.2
Times font as default roman
: S Rahtz]

\begin{document}
\renewcommand{\thefootnote}{\fnsymbol{footnote}}
\sloppy
\newcommand{\rp}{\right)}
\newcommand{\lp}{\left(}
\newcommand \be  {\begin{equation}}
\newcommand \bea {\begin{eqnarray}}
\newcommand \ee  {\end{equation}}
\newcommand \eea {\end{eqnarray}}

\title{Comment on "Are financial crashes predictable?"}
\author{Anders Johansen
\\
The Niels Bohr Institute, University of Copenhagen\\
Blegdamsvej 17, DK-2100 Kbh. \O, Denmark \\
e-mail: johansen@nbi.dk \hspace{5mm}  URL: http//:www.nbi.dk/\~\/johansen
}
\maketitle
\thispagestyle{empty}
\pagenumbering{arabic}

\begin{figure}
%\begin{center}
\parbox[l]{7cm}{
\epsfig{file=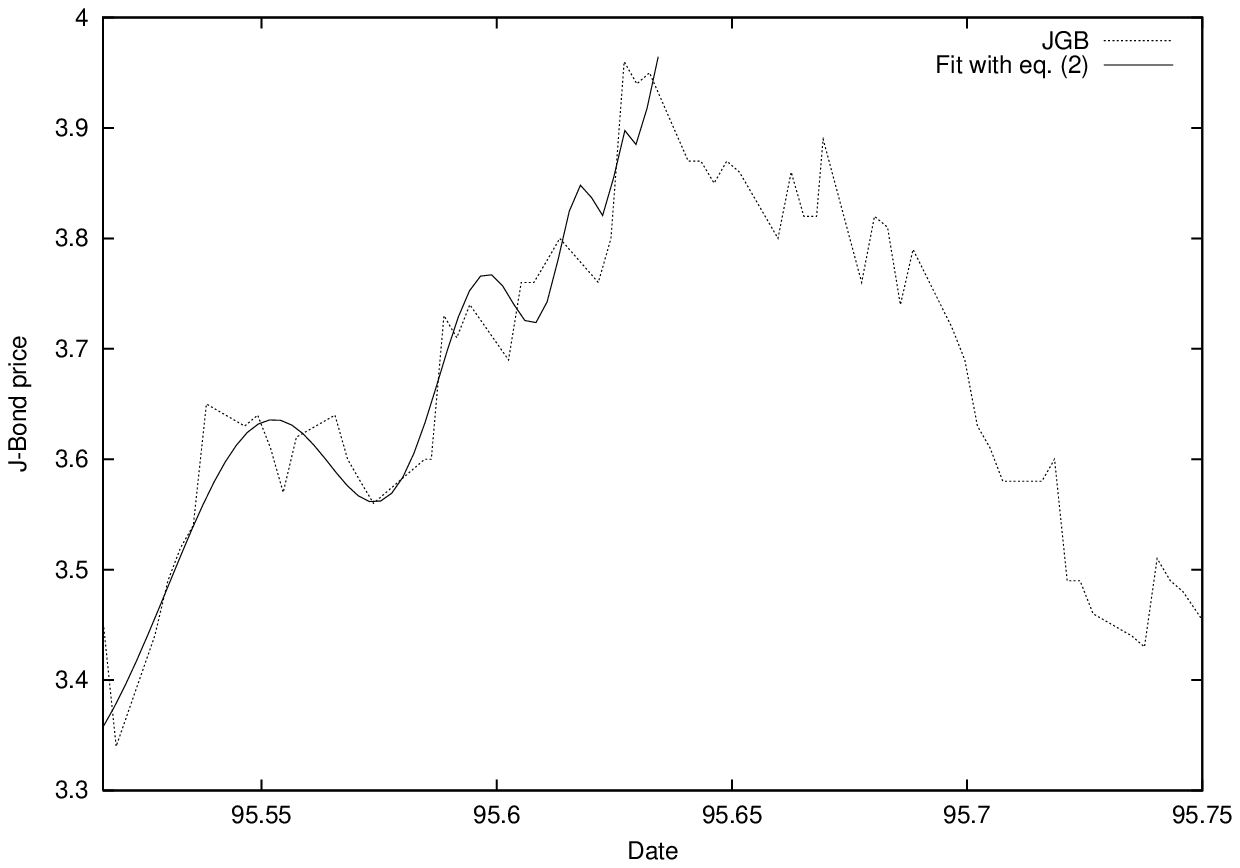,height=7cm,width=8.5cm} 
\caption{\protect\label{lpeqfig} Price of Japanese government bonds fitted
with eq. \protect\ref{lpeq}. The best fit yields $A\approx 3.97$, $B\approx 
-1.95$, $C\approx -0.40$, $t_c\approx 1995.63$, $\beta\approx 0.61$, $\omega
\approx 7.8$ and $\phi\approx 2.8$.}
}\hspace{15mm}
%\end{center}
%\end{figure}
%\begin{figure}
%\begin{center}
\parbox[r]{7cm}{
\epsfig{file=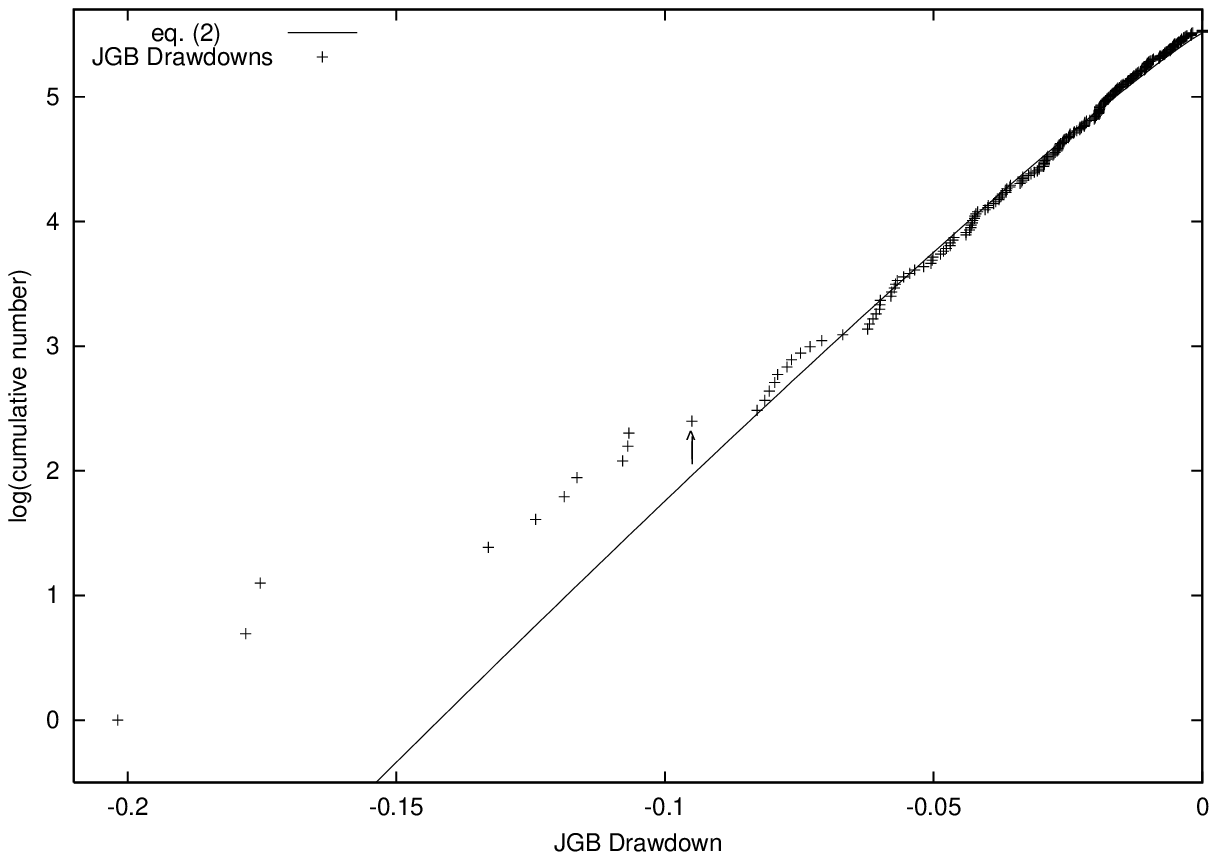,height=7cm,width=8.5cm}
\caption{\protect\label{dddistrib} Drawdown distribution in the price of 
Japanese government bonds 1992.0-1999.2. $\epsilon = 2\%$. The fit with eq. 
(\protect\ref{stretch}) yields $b\approx 46.6$ and $z\approx 1.09$. $a=237$ 
events.}
}
%\end{center}
\end{figure}

In a recent paper published in this journal, Laloux {\it et al.} \cite{Laloux}
criticized the use of eq. 
\be \label{lpeq}
p\lp t\rp = A+B\lp t_c - t\rp^\beta +C\lp t_c - t\rp^\beta \cos\lp  \omega 
\ln\lp t_c - t\rp - \phi \rp 
\ee
as a predictive tool for the detection of periods of large declines in the
financial markets as first suggested in \cite{SJB}. The criticism was based on
a rather primitive ``eye-balling analysis'' lacking the consistent methodology 
used in the identification of more than twenty crashes on the US, Hong-Kong and
FX markets alone which all were preceded by a market bubble parameterised by 
eq. (\ref{lpeq}) \cite{SJ2001}. Specifically, a well-defined optimisation 
algorithm was used fitting eq. (\ref{lpeq}) as well as spectral analysis and 
analysis of synthetic data \cite{SJ2001}. They furthermore wrote ``We want to 
publicly disclose here the fact that on the basis of a log-periodic analysis, a
crash on the JGB market for the end of May 1995 was predicted. On this basis, 
one of us (JPA) bought for $\$1.000.000$ of put options ... The crash did not 
occur and only a delicate trading back allowed to avoid losses''. This 
represents a severely distorted picture of a scientific experiment. First, the 
analysis (performed by the author) and experiment was {\it not} done in May 
1995, but in July 1995. In fig. \ref{lpeqfig} we see eq. (\ref{lpeq}) fitted to
the Japanese bond price in the correct time period of the experiment. Note that
the values obtained for the parameters $\omega$ and $\beta$ (see caption) are 
on the border of what has since been identified prior to over twenty crashes 
on the US, Hong-Kong and FX markets alone \cite{SJ2001}. The result that was 
borne out of the analysis of July 1995 was that JGB might crash around 
$t_c \approx 1995.63 \approx$ 19 Aug. 1995. In fig. \ref{lpeqfig}, we see that
the market peaked on Wednesday 16 Aug. 1995 at a price $3.96$ only to decline 
over $35$ trading days later to a price of $3.36$ on 4 Oct. 1995, {\it i.e.,} 
a total decline of $\approx 15.2\%$. Admittedly, this does not fit the 
conventional picture of a market crash, as it is too slow in its unfolding.
However, considering the lesser volatility generally seen in the bond markets, 
this is not very surprising.

In \cite{outl} it has been shown that a stretched exponential 
\be \label{stretch}
f\lp x\rp = a\exp\lp -bx^z\rp
\ee
is a reasonable null-hypothesis for the bulk of the drawdown distribution of
the FX, major stock markets as well as individual stocks. A drawdown is defined
as a persistent decrease in the price over consecutive days. A drawdown is 
thus the cumulative loss from the last maximum to the next minimum of the 
price. As they are constructed from ``runs'' of the same sign variations
drawdowns thus incorporate higher ($>2$) order correlations not captured
by the distribution of returns. The rationale behind the stretched exponential
is that it constitutes the most obvious extension of the very basic hypothesis 
of a pure exponential. As the definition of ``maximum'' and ``minimum'' is not 
unique except in a strict mathematical sense, a drawdown may be defined in
slightly varying ways. The definition used in the present paper is the 
following. A drawdown is defined as the relative decrease in the price from a 
local maximum to the next local minimum {\it ignoring} relative price increases
in between the two of maximum size $\epsilon$. We will refer to this definition
of drawdowns as ``$\epsilon$-drawdowns'', where we refer to $\epsilon$ as the 
{\it threshold}. In general for stock markets, the largest $<1\%$ events 
did not belong to the same distribution as the bulk parameterised by the 
stretched exponential and have been referred to as ``outliers''. In fig. 
\ref{dddistrib}, the drawdown distribution for JGB the period 1992.00-1999.23 
is shown. We see that the distribution of drawdowns except for the $11 < 5\%$ 
largest events is well captured by eq. (\ref{stretch}). Here we have used
$\epsilon =2\%$ corresponding to the size of a ``normal'' fluctuation.
This result support the existence of outliers in the Japanese bond market. 
That $z > 1$ for JGB further amplifies the suggested existence of outliers 
in the data due to the downward bend in the distribution this implies. The 
arrow on the figure indicates the largest drawdown related to the decline 
ending on 4 Oct. 1995. We see that this drawdown of $9.5\%$ has a clear gap 
of $1.3\%$ to the previous and that it furthermore clearly lies of the fitted
stretched exponential.

In the light of the analysis presented here, the statement in the Laloux 
{\it et al.} article 
\cite{Laloux} re-produced above presents a severely distorted picture of the 
experiment performed in July/Aug. 1995. Not only is the dating of the 
experiment {\it incorrect}, but also the remark that ``only a delicate trading 
back allowed to avoid losses'' is exaggerated considering the actual decline 
in the price. Indeed, the JGB did not crash according to conventional market 
crash definitions as predicted at that time. However, according to the 
non-arbitrary outlier definition advocated by the author and D. Sornette in a 
number of publications \cite{SJ2001,outl} it {\it was} an anomalous event as
shown in fig. \ref{dddistrib}. With respect to the slow unfolding of this 
$15\%$ drop, we note the similarity to the slow stock market crashes of 1962 
and 1998 \cite{SJ2001}. This is also true for the rather large value of 
$\beta$. Last, we want to stress that in order to compare the
use of eq. (\ref{lpeq}) on {\it e.g.}, different times series, the 
``eye-balling
analysis'' used in \cite {Laloux} is completely useless due to the bias such a 
non-objective analysis introduces.

\end{document}